# Artificial skill in monsoon onset prediction: two recent examples

Gerd Bürger[1]
Universität Potsdam, Institut für Umweltwissenschaften und Geographie
Karl-Liebknecht-Str. 24-25, 14476 Potsdam
Germany

**Abstract**

For two cases of empirical monsoon onset prediction it is argued that current verification practice leads to optimistically biased skill, caused by the intricacy of the model setup. For the case of the operational forecasts by the Indian Meteorological Department (IMD) it leads to an overlap of model definition and verification data. A more seriously flawed verification was used in a recent method based on trend extrapolations of 'tipping elements' (TE). Claims of TE of predicting onset 2 weeks earlier than other methods are unjustified. On the contrary, the correlation between TE forecasts and observations is as low as 0.24 and compares poorly to the reported IMD correlation of 0.78. That latter value likely being artificially inflated, currently the best and most reliable monsoon onset predictions come from a dynamical model with more reliable skill values of about 0.7.

**Keywords:** artificial skill; model selection bias; verification; monsoon; forecast

## §1 Introduction

Prediction methods are usually verified by assigning some measure of skill, such as correlation, between forecast and observed values that can be compared to other methods. The skill, of course, should reflect a *real* prediction whose predictand is completely unknown, or at least unknown relative to some reference such as (seasonal) climatology. For empirical-statistical models this can be realized through cross-validation, by withholding one part of the data for the model selection and the other part, the independent data, for verification. This requires stationarity in the data, since the calibrating data should be of the same type as the validating. The advent of statistical computing packages brought automatic validation to current empirical research. Often, however, only the core data-fitting algorithm is cross-validated and not the entire process of model selection. The stationarity and independence requirements are quite strict, nevertheless, so that indeed the reported skills occasionally suffer from a model selection bias or otherwise flawed procedures [*Heckman*, 1977; *Elsner and Schmertmann*, 1994]. The problem is less relevant for dynamical models whose relatively few

1 Corresponding author: Gerd Bürger (*gbuerger@uni-potsdam.de*)

physical (constant) parameters are normally estimated independently of the forecast in question.

How incorrect verification strategies affect the reported skills of empirical models is illustrated by way of two recent studies on the empirical prediction of the monsoon timings. The first one is a study based on which the current operational forecasts by the Indian Meteorological Department (IMD) are issued [*Pai and Nair*, 2009]. The second one is the recent study [*Stolbova et al.*, 2016] who derive a prediction scheme for the Indian summer monsoon from the concept of tipping elements, known from nonlinear dynamical systems theory, hereafter referred to as TE. An earlier critique of TE addressing selection bias has led to a corresponding erratum; further crucial issues remain unresolved, nevertheless, and are addressed here.

## §2 Monsoon onset

Like most monsoonal systems, the Indian summer monsoon is characterized by a sharp transition in the corresponding atmospheric circulation fields and precipitation regimes, so that its onset date (OD) is relatively well defined. Traditionally, monsoon onset over the South-Western Indian state of Kerala (MOK) was defined and recorded by the IMD for more than 100 years. Monsoon advance and retreat over the whole of India is reported by means of isochrone maps, from which OD data can be derived for specific sub-regions. For the MOK, a set of objective criteria (based on rainfall, winds, and outgoing long-wave radiation, cf. [*Joseph et al.*, 2006]) were officially adopted by the IMD in 2006, providing well-defined reference data for OD forecasts. These forecasts are a matter of utmost economic and societal impact. The World Climate Research Programme has established a collaborative structure between the major forecast centers that is solely devoted to improving sub-seasonal to seasonal predictions (cf. [http://www.s2sprediction.net](http://www.s2sprediction.net)), of which monsoon onset prediction is a crucial part. A thorough overview into the considerable amount of corresponding literature is given by [*Goswami and Gouda*, 2010; *Alessandri et al.*, 2015].

### §2.1 Operational IMD forecasts of MOK

To forecast MOK, IMD developed a statistical model based on a principal component regression on a number of atmospheric and surface predictors [*Pai and Nair*, 2009]. At first, using the objective IMD criteria, MOK is defined for each year from 1975 to 2007. Out of an unknown number of potential predictors nine variables are selected for the prediction, based on screening maps of predictand correlations for each year up to 2000, and averaged over the two periods 16-30 April (Model 1) and 1-15 May (Model 2). Given any year from 1997 to 2007, based on the 22 years preceding it the MOK is regressed against the corresponding annual anomalies, and a prediction is made by applying the regression model to that year's anomalies, using Model 1 regressors for the April 30 and Model 2 regressors for the May 15 issue date. For both models, the reported cross-validated correlation with the actual MOK values is 0.78.

### §2.2 Trend extrapolation at tipping elements

By interpreting the monsoon as a dynamical system that alternates between multiple states (basically two, on and off), TE identify OD with the passing of thresholds of some key variables, such as temperature and humidity, at certain tipping elements. The tipping elements idea has been introduced into climate science a while ago [*Held and Kleinen*, 2004; *Lenton et al.*, 2008], and it can ultimately be traced back to earlier work on order parameters and self organization from the physics and chemistry literature [*Haken*, 1978; *Ginzburg and Landau*, 2009]. As a precursor to passing such a threshold a growth of fluctuations in those variables is often observed, which is used by TE to identify the region of the Eastern Ghats (EG) and North Pakistan (NP) as the main tipping elements. The prediction scheme uses linear temperature trend extrapolations at one tipping element, $T_{NP}$, and estimates OD as the passing dates of the corresponding climatological thresholds at the other, $T_{EG}$: $T_{NP}(t) > \langle T_{EG}(\langle OD \rangle) \rangle$. This configuration (the role of EG and NP and the selection of T) was chosen as the one that is least susceptible to estimation error. Verification is done against the OD values for EG, as estimated from NCEP reanalysis fields, allowing an uncertainty range of 7 days (from the supplement, cf. [*Singh and Ranade*, 2010]). This results in a success rate of 73% for predictions issued on May 5 of each year. In comparison to that (as stated in the TE supplement), the constant predictor (climatology) achieves a success rate of 75%. According to TE, this drop of 2% 'reflects the distribution of the monsoon onset dates over the long-term period', which seems to express the trivial fact that variable targets are harder to predict than static ones.

## §3 Critique

### §3.1 Model selection bias

In general, model selection bias describes a case when estimates of prediction skill are positively biased ('artificial') due to inadequate sampling, specifically, if model building and model verification are based on overlapping data. In short: if prior to its final definition the model is probed for the resulting skill. Figure 1 shows a scheme to illustrate model skill S and bias in the most simple case of a single model parameter p. Given that in real world cases the estimated model skill Ŝ depends on imperfectly sampled data, so that $Ŝ(p) \neq S(p)$, the goal is twofold: a) estimate $p_{opt}$ that maximizes the true skill function, $S(p)$, and b): estimate $S(p_{opt})$. If, for a given sample, one chooses p that maximizes $Ŝ(p)$ for that sample, $Ŝ(p̂) = \max_p(Ŝ(p))$, one will most certainly obtain one for which $Ŝ(p̂) \neq S(p_{opt})$. Except for some rare cases, the estimate p̂ will be unbiased, $\langle p̂ \rangle = p_{opt}$. However, as evident from the figure this does not extend to the skill itself: $\langle Ŝ(p̂) \rangle > S(p_{opt})$, the latter equation describing the model selection bias.

Unless the parameter p represents overfitting, that is, its best estimate is a sample property and does not generalize well, the model skill is a smooth function of p as in the figure, thus representing the limiting function of its in-sample estimates. Conversely, unless the data are nonstationary and no empirical model should be build anyway, the best skill estimate from the full sample forms an approximate upper bound of the true skill.

### §3.2 IMD

The screening of predictors is based on correlation maps from the period 1975-2000. Hence there is an overlap of 4 years, or 36% of data, with the verification period 1997-2007. This is aggravated by the relatively strong interannual autocorrelations in that area, which further increases the overlap. Here the set of potential predictors plays the role of the model parameter p in Figure 1. Similar artificial skill, potentially with a larger overlap, was observed by [*DelSole and Shukla*, 2009] for the operational forecasts of monsoon intensity [*Rajeevan et al.*, 2007].

[*Pai and Nair*, 2009] do not report to predict a MOK centered about its long-term mean (June 1), so I will assume they do not. If otherwise that mean is estimated from the full dataset, additional artificial skill has to be factored in from that extra overlap.

### §3.3 TE

Using the EG region instead of Kerala and NCEP-based OD as a reference may be a potential new route to empirical monsoon forecasting. The verification employed by TE, however, contains serious methodological flaws. Obviously, when offering alternatives to existing methods and declaring them as superior, verification is paramount. After stating that the constant predictor (i. e. climatology) has superior skill one would normally conclude that the prediction is of little use. But even within the lines of argument that lead to this statement at least three flaws are identified:

a) allowing for a *predictand* uncertainty

b) 'cherry picking' the period 2005-2015 for verification

c) incorrect assessment of competing methods (lead times and p-values)

ad a)   While probabilistic *predictors* are fairly common in various scientific branches, probabilistic or uncertain *predictands* would imply that the outcome of an experiment or observation is still unknown even post factum. This is fairly uncommon and has, to my knowledge, never been used. It definitely requires that a corresponding prediction cannot be more skillful if that uncertainty becomes larger. For example, TE allows for a range of 7 days in predicting the OD. That number pertains to the uncertainty of the NCEP-estimate and not the recorded OD. Following that logic, one would only have to look for an OD-estimate worse than NCEP to easily obtain higher skill ('hedge' the prediction).

ad b)   A major critique of the original (pre-erratum) TE paper was a too-optimistic verification due to selection bias; no distinction was made between model definition and verification, leading to a 100% overlap of corresponding data. But even the in-sample correlation skill was poor, with only $\rho=0.24$ and a p-value of 0.09, as shown in Figure 2 with my own digitized data. Following the arguments of §3.1, this renders the model basically useless, which is also in accordance with the insignificant success rate mentioned earlier. But TE try to circumvent the selection bias anyway, by employing a single data-split into 1965-2004 for calibration and 2005-2015 for validation, and find for the latter a correlation of $\rho=0.62$ (p=0.04, cf. Table S3 of TE). Whatever the argument for picking that particular validation period, it contradicts the stationarity assumption for crossvalidation and for building the empirical model in the first place. For example, if the pre-2005 data are bad, how can the model be calibrated on those? - And even then, since the predictand is NCEP based and applies to EG and not Kerala, no reason exists for preferring one period to another.

ad c)   The competing methods are summarized in Table S3 of TE; it is derived from a table that was produced for the review of the TE erratum and which is reproduced here as Table 1 for comparison. Model 1 forecasts by IMD, issued on April 30, have a reported skill of 0.78 but were removed for Table S3. As regards to p-values, they are a function of correlation and sample size (and do not need to be reported separately), so that a reported correlation of 0.78 for both IMD predictions [Pai and Nair, 2009, Model 1 and 2] implies a p-value of 0.002 and thus a very high significance. With regard to forecast range (Table S3, col. 5), the IMD forecast issue times are primarily the result of customer demand than of scientific endeavor, which should not be used to discount the corresponding forecast. For [Alessandri et al., 2015] only the earlier-than-normal ODs are considered, which of course reduces the resulting forecast range in comparison to TE; the appropriate range is 30 days.

Consequently, if it has any skill at all (no significant skill at $\alpha=5\%$), the TE forecast skill of $\rho=0.24$ compares quite poorly to the OD forecasts for Kerala by IMD, issued on April 30 and based on the period 1997-2007, with a correlation of $\rho=0.78$ of [Pai and Nair, 2009], and to those of [Alessandri et al., 2015] with $\rho=0.70$ (issued on May 1 and based on 1989-2005). For this particular case, hence, the existing methods prove to be largely superior in forecasting the OD.

### §4 Conclusions

Several flaws were uncovered in the verification procedures of two empirical monsoon prediction schemes. Both schemes, the operational IMD forecasts as well as the recent approach by [*Stolbova et al.*, 2016] using tipping elements (TE), suffer from incorrect verification, so that the reported skills are overly optimistic. The IMD model suffers from a model selection bias with a 36% overlap of model definition and verification data; this should not be too difficult to correct, and the overlap is moderate enough for leaving the skill roughly as reported. For TE, if its full dataset is sufficiently stationary to allow calibration and validation of an empirical model in the first place, the true skill will not be much different from the in-sample estimate of $\rho=0.24$, and is worse than climatology in terms of success rates (73% vs. 75%). The reported higher skill ($\rho=0.62$) for the particular choice of a calibration/validation split is based on a similar selection bias and therefore invalid. No significantly nonzero predictive skill is obtained for the TE forecast scheme.

In light of this, a comprehensive monsoon prediction inter-comparison effort is needed that follows a strict verification protocol. It should include the clear definition of a predictand, preferably based on objective criteria (similar to the MOK), and ensure complete independence of the verification dataset. The Subseasonal-to-seasonal (S2S) prediction project (cf. *http://s2sprediction.net*) of the WMO provides a perfect umbrella for such an undertaking. Until then, the currently best verified prediction method for monsoon onset appears to be the dynamical forecasts by [*Alessandri et al.*, 2015].

A final word on falsification or, more generally, on falsifiability. Disproving (falsifying) theories or facts has always been at the heart of the scientific endeavor, and according to [*Popper*, 1935] falsifiability is indeed its defining element. In times of ever growing data to be analyzed and understood, with an equally growing market of empirical-statistical schemes to do the job, and in times when obtaining a scientific degree indispensably requires to publish (and to publish more than once), the delicate balance between 'proving' and disproving facts and theories is endangered, as the tide inevitably turns towards the former. It may be time to remind ourselves, hence, that the critical inspection of a scientific result is no less a result than the result inspected, helping to restore science to the level of due diligence that Popper had in mind.

### Acknowledgments

Helpful comments by two anonymous reviewers of my original TE criticisms are greatly appreciated. No data sharing issues apply since all of the numerical information is contained in the cited figures. Despite multiple requests, no data were provided by [*Stolbova et al.*, 2016].

Digitizing was done using the PlotDigitizer program (*http://plotdigitizer.sourceforge.net*), which works manually and is thus subject to normal and random errors; the digitized data are archived at https://www.researchgate.net/profile/Gerd_Buerger.

| method issue date | period | reference target area | correlation | no-skill p-value [%] | source |
|---|---|---|---|---|---|
| **IMD Apr 30** | **1997-2007** | **IMD Kerala** | **0.78** | **0.2** | **Pai and Nair, 2009** |
| **Alessandri May 1** | **1989-2005** | **IMD Kerala** | **0.70** | **0.08** | **Alessandri et al., 2015** |
| TE May 5 | 1965-2015 | NCEP EG | 0.28 | 2.5 | TE, digitized |
| TE May 5 | 2005-2015 | NCEP EG | 0.64 | 1.7 | TE, digitized |
| TE May 5 | 1965-2007 | IMD EG | 0.24 | 6.2 | Singh and Ranade, 2010, digitized TE, digitized |

Table 1. Monsoon onset date forecast verification. Skillful forecasts bold ($\alpha=1\%$).

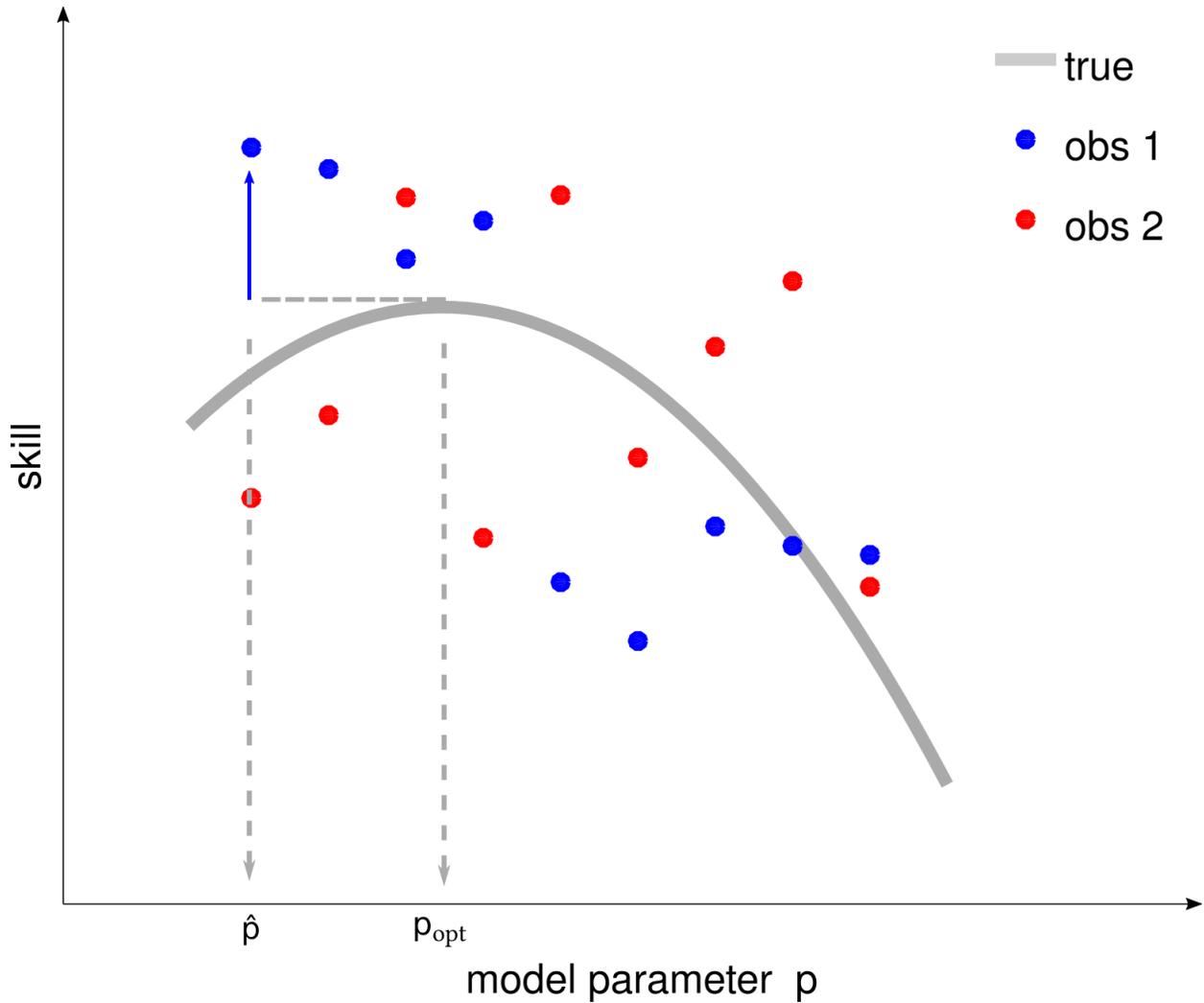

*Figure 1. Schematic exemplifying model selection bias. Gray line: The true but unobserved skill depending on model parameter p, S(p). Goal: obtain best estimate of $p_{opt}$ and $S(p_{opt})$ so that $S(p_{opt})$ is optimal. Blue: Model selection based on maximum skill at $\hat{p}$ for observational sample 1 (obs 1), the corresponding selection bias being indicated by the arrow: $\langle \hat{S}(\hat{p}) \rangle > S(p_{opt})$. New data (obs 2, red) from a real prediction will generally have less skill.*

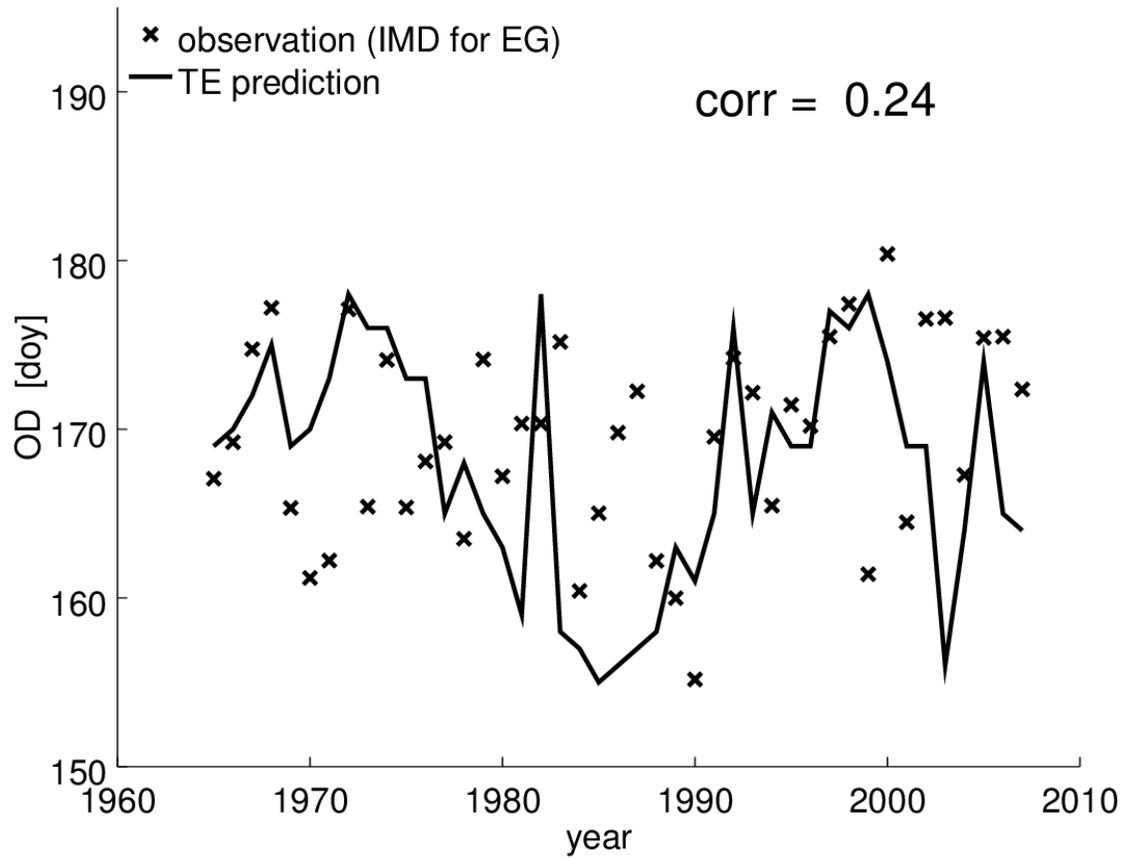

*Figure 2. Annual monsoon onset date (OD) for the EG, as determined by IMD (crosses) and as predicted by TE (line). Data based on digitization of [Singh and Ranade, 2010], Figure 20 and TE, Figure 4. The correlation of 0.24 is not significantly nonzero (p-value = 0.12).*